\documentclass[prl,twocolumn,aps,epsf]{revtex4}
\usepackage{amssymb}
\usepackage{amsmath}
\usepackage{epsfig}

\input{tcilatex}

\begin{document}

\title{Field-induced superconductivity with enhanced and tunable
paramagnetic limit}
\author{A. Buzdin$^{\ast }$, S. Tollis, and J. Cayssol}
\affiliation{Condensed Matter Theory Group, CPMOH, UMR 5798, Universit\'{e} Bordeaux I,\\
33405 Talence, France}
\affiliation{(*)also Institut Universitaire de France, Paris}

\begin{abstract}
We demonstrate that in a superconducting multilayered system with
alternating interlayer coupling a new type of \ nonuniform superconducting
state can be realized under in-plane magnetic field. The Zeeman effect in
this state is compensated by the energy splitting between bonding and
antibonding levels. Such compensation mechanism at low temperature leads to
the field-induced superconductivity. We discuss the conditions for the
experimental observation of the predicted phenomena.
\end{abstract}

\maketitle

\bigskip There are two mechanisms of the superconductivity destruction by a
magnetic field: orbital and paramagnetic effects \cite{AGD,James}. Usually
it is the orbital effect that is more restrictive. However in the systems
with large effective mass of electrons \cite{heavy1,heavy2} or in
low-dimensional compounds, like quasi-one-dimensional or layered
superconductors under in-plane magnetic field \cite{uji}, the orbital
magnetism is weakened and it is the paramagnetic effect which is responsible
for the superconductivity destruction. The Chandrasekhar-Clogston
paramagnetic limit \cite{chandra,clogston} is achieved when the energy of
the polarization of the normal electron gas, $-\chi _{n}H^{2}/2,$ equals the
superconducting condensation energy $-N(0)\Delta _{0}^{2}/2$ , where $N(0)$
is the density of states of the normal electron gas, $\chi _{n}$ its spin
susceptibility and $\Delta _{0\text{ }}=1.76T_{c}$ is the zero temperature
superconducting gap. This gives the critical field $H_{p}=\Delta _{0}/(\sqrt{%
2}\mu _{B})$ of the first order transition at $T=0$, $\mu _{B}$ being the
Bohr magneton. Later Larkin and Ovchinnikov \cite{lo} and Fulde and Ferrell 
\cite{ff} (FFLO) predicted the appearance at low temperature of a nonuniform
superconducting state with the zero temperature critical field $%
H_{3D}^{FFLO}=0.755\Delta _{0}/\mu _{B}$,\textit{\ i.e.} somewhat higher
than the paramagnetic limit $H_{p}$. This prediction was made for
three-dimensional\ superconductors. In quasi-two-dimensional superconductors
the critical field of the FFLO state is even higher, namely $%
H_{2D}^{FFLO}=\Delta _{0}/\mu _{B}$ \cite{bulaevski73}, while in
quasi-one-dimensional systems there is no paramagnetic limit at all \cite%
{buzdin83}. The appearance of the modulated FFLO state is related to the
pairing of electrons with opposite spins which do not have the opposite
momenta anymore due to the Zeeman splitting.

In this Letter, we demonstrate that in a ballistic superconducting bilayer
at low temperature and strong enough coupling $t_{1}\gg \Delta _{0}$ between
the conducting planes, the paramagnetic limit is enhanced up to $H_{c}\sim
t_{1}/\mu _{B}$ far above the usual limit $H_{2D}^{FFLO}=\Delta _{0}/\mu
_{B} $. More precisely, a very unusual superconducting phase is settled
between a lower and an upper critical fields given by $\mu
_{B}H_{c}=t_{1}\pm \mu _{0}^{2}H_{0}^{2}/t_{1\text{ }}$and below a maximal
temperature of the order of $T_{c}^{2}/t_{1}$. Thus one obtains \textit{%
field induced superconductivity} above the lower field while the upper one
provides\textit{\ the enhanced paramagnetic limit which may be tuned by
varying the electronic coupling} $t_{1}$. This is due to the compensation of
the the Zeeman splitting by the energy splitting $t_{1}$ between bonding and
antibonding electronic states of the bilayer. As another important feature
of this new phase, adjacent layers support opposite signs of the order
parameter. Note that such a so-called $\pi $ phase was predicted before \cite%
{andreev91,houzet02} for the superconductor-ferromagnet multilayered systems
where the atomic superconducting and ferromagnetic layers alternated and
were weakly coupled in contrast to the present system.

\begin{figure}[tbph]
\includegraphics[height=5.2cm]{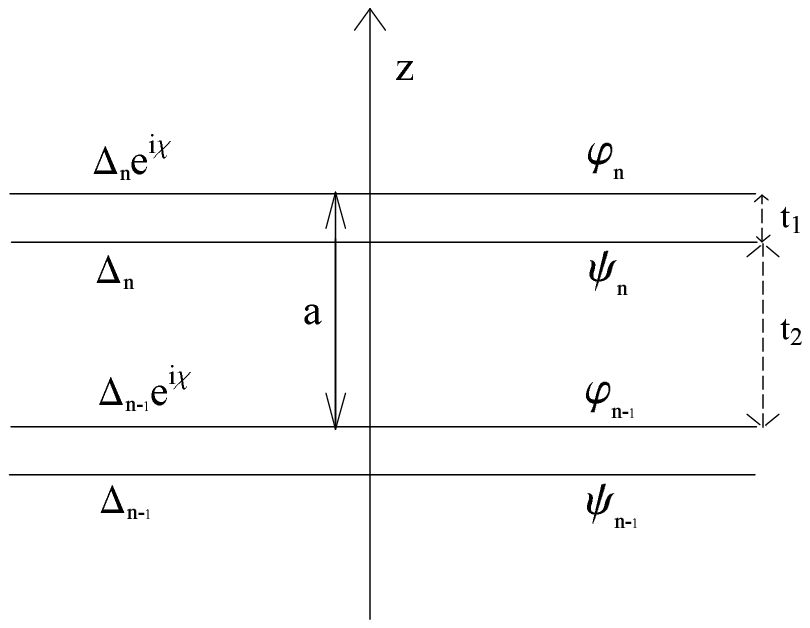}
\caption{{Multilayered system. The n$^{th}$ unit cell contains two
superconducting planes $\protect\varphi _{n}$ and $\protect\psi _{n}$. The
transfer integrals are very different, $t_{2}\ll t_{1}$. The superconducting
phase difference \ $\protect\chi $ between two adjacent planes can be either 
$0$ or $\protect\pi $.}}
\end{figure}

We consider a model multilayered system with a cristallographic structure
similar to those of the high-$T_{c\text{ }}$superconductors \cite{bulaevki90}%
, as shown in Fig. 1. Namely we assume that the electrons are confined in
the atomic planes with the same zero-field dispersion relation \ $\xi (%
\mathbf{p})=\mathbf{p}^{2}/2m-E_{F}$, $\ E_{F\text{ }}$being the Fermi
energy. The transfert integrals between the planes are different $t_{1}\gg $ 
$t_{2}$, both being smaller than the Debye energy $\hbar \omega _{D}\ll
E_{F} $ . In this structure the $n^{th}$ unit cell contains two conducting
planes, labelled by $\psi _{n}$ and $\varphi _{n}$. The coordinate in the
plane is\textbf{\ }$\mathbf{r}$ and the cells are separated by a distance $a$
along the $z$-axis which is chosen perpendicular to the planes. We suppose
that the Cooper pairing occurs in the planes and perform our analysis in the
framework of the standard mean field BCS Hamiltonian,

\begin{equation}
H=H_{\psi 0}+H_{\varphi 0}+H_{t}+\frac{1}{\left\vert \lambda \right\vert }%
\int \mathbf{d}^{2}\mathbf{r}\Delta _{n}^{2}(\mathbf{r}),
\label{hamiltonian}
\end{equation}%
with 
\begin{eqnarray}
H_{\psi 0} &=&\underset{\mathbf{p},n}{\sum }\xi (\mathbf{p})\psi _{n,\alpha 
\text{ }}^{\dagger }(\mathbf{p})\psi _{n,\alpha \text{ }}^{{}}(\mathbf{p}) 
\notag \\
&&+\frac{1}{2}\Delta _{n,\alpha \beta }(\mathbf{q})\psi _{n,\alpha \text{ }}(%
\mathbf{p}+\mathbf{q})\psi _{n,\beta \text{ }}(-\mathbf{p})+h.c.,  \notag \\
H_{t} &=&\underset{p}{\sum }\psi _{n,\alpha \text{ }}^{\dagger }(\mathbf{p}%
)\left( t_{1}\varphi _{n,\alpha \text{ }}^{{}}(\mathbf{p})+t_{2}\psi
_{n+1,\alpha \text{ }}^{{}}(\mathbf{p})\right) +h.c.,  \notag
\end{eqnarray}%
where $\lambda $ is the BCS coupling constant, summation over repeated spin
indexes $\alpha ,\beta $ is implied, and $\Delta _{n,\alpha \beta }(\mathbf{q%
})$ is the Fourier transform of the superconducting order parameter $\Delta
_{n,\alpha \beta }(\mathbf{r})=\Delta _{n}(\mathbf{r})i\sigma _{\alpha \beta
}^{y}$ , $\sigma ^{y}$ being the second Pauli matrix. The operators $\psi
_{n,\alpha \text{ }}(\mathbf{p})$ and $\varphi _{n,\alpha \text{ }}(\mathbf{p%
})$ destroy one electron with spin $\alpha $ and momentum $\mathbf{p}$,
respectively in planes $\psi _{n}$ and $\varphi _{n}$. Our model includes
both translationnal and jauge symmetry breaking. Indeed, the superconducting
order parameter in the planes $\psi _{n}$ and $\varphi _{n}$ \ are
respectively given by $\Delta _{n}(\mathbf{r})=\Delta e^{i\mathbf{q}.\mathbf{%
r}+i\kappa na}$ and $\Delta _{n}(\mathbf{r})e^{i\chi }$, where $\mathbf{q}$
and $\kappa $ are respectively the in-plane and the perpendicular modulation
wave vectors. These vectors and the superconducting phase difference $\chi $
must be determined from the minimum energy condition.

The Gor'kov equations corresponding to the Hamiltonian (\ref{hamiltonian})
are solved exactly. In the small $\Delta $ limit, the linearized anomalous
Gor'kov Green function $F_{\omega }^{\dagger }=\left\langle \psi
_{\downarrow \text{ }}^{\dagger }(\mathbf{p})\psi _{\uparrow \text{ }%
}^{\dagger }(-\mathbf{p})\right\rangle $ reduces to the form \cite{footnote1}

\begin{equation}
F_{\omega }^{\dagger }=\frac{\Delta (t_{k}^{\ast }\widetilde{t_{k}}e^{-i\chi
}-\widetilde{\omega }_{-}\omega _{+})}{(\widetilde{\omega }%
_{-}^{2}-\left\vert \widetilde{t_{k}}\right\vert ^{2})(\omega
_{+}^{2}-\left\vert t_{k}\right\vert ^{2})},  \label{propagator}
\end{equation}%
where $t_{k}=t_{1}+t_{2}e^{ika}$, $\omega _{\pm }(\mathbf{p})=i\omega \pm
\xi (\mathbf{p})-\mu _{B}H$, $\widetilde{t}=t_{k+\kappa }$ and $\tilde{\omega%
}_{\pm }(\mathbf{p})=$ $\omega _{\pm }(\mathbf{p}+\mathbf{q})$, $\omega $
being the Matsubara frequencies.

We first solve the isolated bilayer problem $t_{2}=0$. After integration of
Eq.(\ref{propagator}) over $\xi (\mathbf{p})$, one obtains the anomalous
Eilenberger propagator $f_{\omega }^{\dagger }(\mathbf{v}_{F})=\int (d\xi
/\pi )F_{\omega }^{\dagger }=\sum_{a=\pm 1}f_{\omega ,a}^{\dagger }$ with 
\begin{equation}
\frac{f_{\omega ,a}^{\dagger }}{\Delta }=\frac{\omega +i\mu _{B}H+i\frac{%
\mathbf{v}_{F}\cdot \mathbf{q}}{2}+iat_{1}e^{-i\chi /2}\cos (\chi /2)}{%
2(\omega +i\mu _{B}H+i\frac{\mathbf{v}_{F}\cdot \mathbf{q}}{2})(\omega +i\mu
_{B}H+i\frac{\mathbf{v}_{F}\cdot \mathbf{q}}{2}+iat_{1})}
\label{xiintegrated}
\end{equation}%
for positive $\omega $, where $a=\pm 1$ labels bonding and antibonding
states while $\mathbf{v}_{F}$ is the Fermi velocity vector in the plane.
Then the self-consistency relation (\ref{self}) below implies that the
superconducting phase difference $\chi $ between neighboring layers is
either zero or $\pi $. In the absence of Zeeman splitting $H=0$, the
superconducting order parameter is naturally the same in both layers, namely 
$\chi =0$. In this case, $f_{\omega }^{\dagger }=\Delta /\omega $ coincides
with the well-known anomalous Green function of a superconductor in the
limit $\Delta \rightarrow 0$, and the self-consistency relation gives the
bare critical temperature $T_{c}$ of the isolated layer. More generally for $%
\chi =0$ and finite $H$, the interlayer coupling $t_{1}$ drops from Eq.(\ref%
{xiintegrated}) and one retrieves $f_{\omega }^{\dagger }=\Delta /(\omega
+i\mu _{B}H)$\ for a two dimensional superconductor under parallel magnetic
field. The other possible choice is $\chi =\pi $, when the superconducting
order parameter is opposite on adjacent layers. For small values of the
magnetic field, this later $\pi $ phase exhibits naturally a lower critical
temperature than the $\chi =0$ phase.

\begin{figure}[tbph]
\includegraphics[scale=0.43]{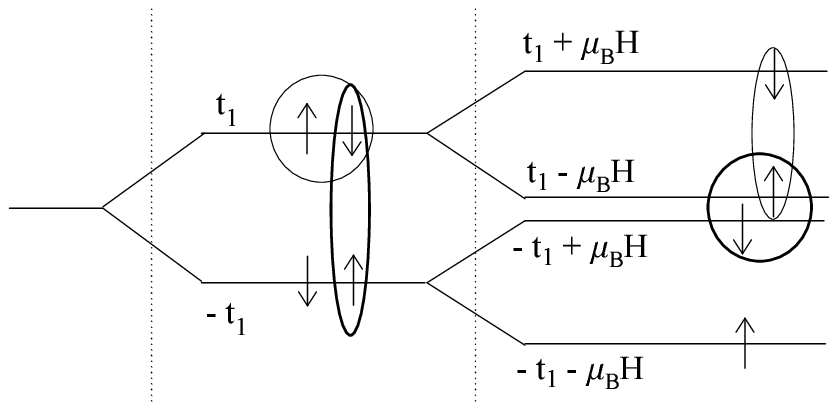}
\caption{Excitation spectrum. Usual singlet pairing (thin line circles)
between opposite-spin electrons occupying the same orbital is affected by
Zeeman effect. In contrast, $\protect\pi $ coupling (thick line) between two
electrons occupying a bonding and an antibonding orbitals may lead to the
cancellation of the Zeeman splitting.}
\end{figure}

However for relatively large interlayer coupling $t_{1}>T_{c}$ and high
field, the situation becomes drastically different. Indeed, the excitation
spectrum consists of four different branches $\epsilon =\pm \xi +\mu
_{B}H\pm t_{1}$ in the limit $\Delta \rightarrow 0$, see Fig. 2. The singlet
pairing may occur here between \textit{one electron in the bonding orbital
and the other electron in the antibonding orbital}. This results in a very
special coupling where, if $\mu _{B}H=t_{1}$, the Zeeman splitting is
exactly compensated. Therefore enhanced superconductivity is expected in the
vicinity of $\mu _{B}H=t_{1}$, at least at zero temperature.

In order to derive rigourously this prediction, we analyse the
self-consistency relation, 
\begin{equation}
\Delta =2\pi \left\vert \lambda \right\vert N_{2D}(0)T\underset{}{\underset{%
\omega >0,a}{\sum }\func{Re}\left\langle f_{\omega ,a}^{\dagger }(\mathbf{v}%
_{F})\right\rangle },  \label{self}
\end{equation}%
in the $\pi $ phase where $f_{a,\omega }^{+}(\mathbf{v}_{F})$ $\ $depends on
the coupling $t_{1}$ in the following way

\begin{equation}
f_{\omega ,a}^{\dagger }(\mathbf{v}_{F})=\frac{\Delta }{2(\omega +i\mu
_{B}H+iat_{1}+i\mathbf{v}_{F}\cdot \mathbf{q}/2)}.  \label{anomalouspi}
\end{equation}%
Here $N_{2D}(0)=m/(2\pi )$ is the two-dimensional density of states per unit
surface and per one spin orientation, and the brackets $\left\langle
...\right\rangle $ denotes averaging over the polar angle $\theta =(\mathbf{v%
}_{F},\mathbf{q})$ .

We first discuss the zero temperature second-order phase transition between
the normal metal and the $\pi $ phase, as a function of the magnetic field.
From Eqs.(\ref{self},\ref{anomalouspi}), the critical field $H$ is shown to
satisfy

\begin{multline}
\left\vert H+t_{1}/\mu _{B}+\sqrt{(H+t_{1}/\mu _{B})^{2}-X^{2}}\right\vert
\label{HCritFFLO} \\
.\left\vert H-t_{1}/\mu _{B}+\sqrt{(H-t_{1}/\mu _{B})^{2}-X^{2}}\right\vert
=4H_{0}^{2}\text{ ,}
\end{multline}%
where $X=\left\vert \mathbf{q}\right\vert v_{F}/(2\mu _{B})$, and $%
H_{0}=\Delta _{0}/2\mu _{B}$ \ is the critical field of the second-order
superconducting phase transition in a two-dimensional monolayer. One must
then find the value of $X$ which maximizes the critical field $H$. If \ the $%
\pi $ phase is assumed to be uniform inside each plane, namely if $\mathbf{q}%
=\mathbf{0}$, Eq.(\ref{HCritFFLO}) \ merely reduces to \ $\left\vert
H^{2}-t_{1}^{2}/\mu _{B}^{2}\right\vert =H_{0}^{2}$ and we obtain a lower
and an upper critical fields respectively given by $\mu _{B}H_{c}=t_{1}\pm
\mu _{B}^{2}H_{0}^{2}/2t_{1}$, in the limit $t_{1}\gg \mu _{B}H_{0}$. Thus
at zero temperature and strong enough coupling $t_{1}\gg \Delta _{0}$, the
superconductivity destruction follows a very special scenario. At low
fields, superconductivity is first suppressed in the usual manner at the
paramagnetic limit $H_{2D}^{FFLO}=\Delta _{0}/\mu _{B}$ leading to the
normal metal phase. Then further increase of the field leads to a normal to
superconducting phase transition at the lower critical field. This
superconducting $\pi $ phase is finally suppressed at the upper critical
field. This is a new paramagnetic limit which may be tuned far above the
usual one merely by choosing the coupling $t_{1}$ greater than $\Delta _{0}$%
. Thorough analysis of \ Eq.(\ref{HCritFFLO}) shows that the upper critical
field is even increased by an in-plane modulation in analogy with the
two-dimensional FFLO phase \cite{bulaevski73}. In this sense, the
low-temperature $\pi $ phase in the present case should be called a FFLO-$%
\pi $ phase. The upper critical field is maximal for the choice $%
X=\left\vert \mathbf{q}\right\vert v_{F}/2\mu _{B}=\left\vert H-t_{1}/\mu
_{B}\right\vert $, and then Eq.(\ref{HCritFFLO}) reduces to 
\begin{equation}
\left\vert H-t_{1}/\mu _{B}\right\vert .\left\vert H+t_{1}/\mu _{B}+2\sqrt{%
Ht_{1}/\mu _{B}}\right\vert =4H_{0}^{2},  \label{Hupperlower}
\end{equation}%
which gives the upper and lower fields $\mu _{B}H_{c}=t_{1}\pm \mu
_{B}^{2}H_{0}^{2}/t_{1}$ in the $t_{1}\gg \Delta _{0}$ limit. Note that the
period of the modulated order parameter $\left\vert \mathbf{q}\right\vert
^{-1}=\xi _{0}(t_{1}/\Delta _{0})$ is larger than the corresponding period
in the two-dimensional FFLO phase which coincides with the ballistic
coherence length $\xi _{0}=v_{F}/\Delta _{0}$ \cite{bulaevski73}.

\begin{figure}[tbph]
\includegraphics[scale=0.43]{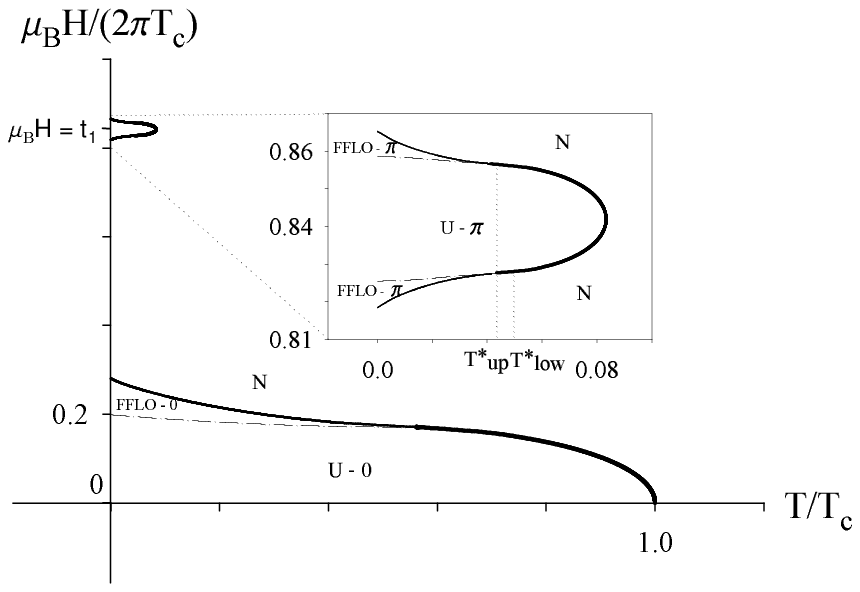}
\caption{Phase diagram for $t_{1}=3\Delta _{0}$. U-$\protect\chi $ (resp.
FFLO-$\protect\chi $) denotes the uniform (resp. modulated) superconducting
state with phase difference $\protect\chi $ between the planes. Thick (resp.
thin) solid lines represents second-order transition between U-$\protect\chi 
$ (resp. FFLO-$\protect\chi $) and normal metal phase (N). We expect the U-$%
\protect\chi $/ FFLO-$\protect\chi $ transition lines (not calculated) to be
in the vicinity of the (virtual) first order U-$\protect\chi $/normal metal
lines (dash-dotted).}
\end{figure}

Furthermore one may derive the full temperature-field phase diagram using
Eqs.(\ref{self},\ref{anomalouspi}) and the result is shown in Fig. 3. When
the temperature is increased, the lower critical field increases whereas the
upper one decreases. Along the upper (resp. lower) critical line the FFLO
modulation is lost at some temperature $T_{up}^{\ast }$ (resp. $%
T_{low}^{\ast }$). For higher temperatures a uniform $\pi $ phase (U-$\pi $)
is recovered and the temperature dependence of the critical field is given by

\begin{equation}
\ln \frac{T}{T_{c}}=\underset{a=\pm 1}{\frac{1}{2}\sum }\func{Re}\left[ \Psi
\left( \frac{1}{2}\right) -\Psi \left( \frac{1}{2}+i^{{}}\frac{\mu
_{B}H_{c}(T)+at_{1}}{2\pi T}\right) \right] ,  \label{HcriticalTemperature}
\end{equation}%
where $\Psi \left( x\right) $ is the Digamma function and $\Psi
(1/2)=-C-2\ln 2\simeq -1.963$, $C$ being the Euler constant. Finally the
lower and upper critical lines merge at field $H_{c}=t_{1}/\mu _{B}$ and
temperature $T_{M}=\pi e^{-C}T_{c}^{2}/(4t_{1})$ in the limit $t_{1}\gg T$.
Therefore the field induced $\pi $ superconductivity is confined to
temperatures lower than $T_{M}$. The structure of these U-$\pi $ and the
FFLO-$\pi $ phases is reminiscent of the corresponding U-$0$ and the FFLO-$0$
phases although the former are shifted to higher fields and lower
temperatures than the later. 
\begin{figure}[tbph]
\includegraphics[height=4.9cm]{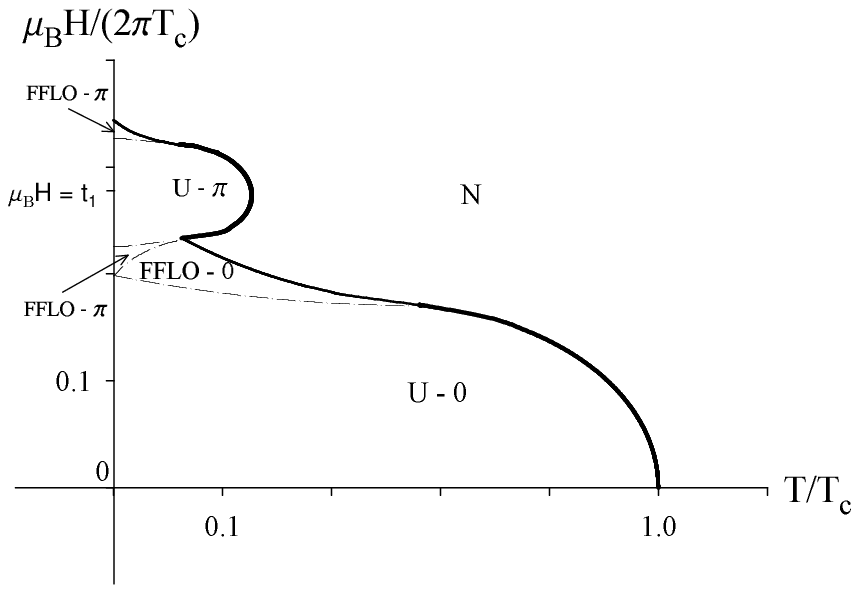}
\caption{Phase diagram for $t_{1}=\Delta _{0}$. Solid and dash-dotted lines
have the same meaning than in Fig. 3.}
\end{figure}

Above results were obtained for relatively strong coupling. For lower
coupling $t_{1}\simeq \Delta _{0}$, the U-$\pi $ and the FFLO-$\pi $ phases
merge continuously into the usual $\chi =0$ phases as shown in Fig. 4, and
finally disappear for $t_{1}$ slightly smaller than $\Delta _{0}$. From an
experimental point of view, one might choose a system with intermediate
coupling $t_{1}$ small enough to settle the $\pi $ phase island in an
available range of temperatures but also large enough to separate the $\pi $
phase island from the usual superconducting phases with $\chi =0$.

\bigskip

Hereafter we discuss various physical mechanisms limiting the above
predicted $\pi $ superconductivity: role of finite $t_{2}$, impurity and
orbital effects.

For finite coupling $t_{2}$ between the bilayers, the self-consistency
relation (\ref{self}) together with Eq.(\ref{propagator}) leads to the
following equation

\begin{equation}
\left\langle \ln \frac{2\mu _{B}H_{0}}{\underset{a=\pm 1}{\sum }\sqrt{(\mu
_{B}H+X\cos \theta )^{2}-(t_{1}+at_{2})^{2}}}\right\rangle =0.
\label{t2effet}
\end{equation}%
For $X=\left\vert H-(t_{1}+t_{2})/\mu _{B}\right\vert $, the singularity
around $\theta =\pi $ produces a corrective term to the upper and lower
critical fields found previously. However this correction is negligible if $%
t_{1}t_{2}\ll \mu _{B}^{2}H_{0}^{2}$. In the opposite regime, that is for
larger values of the inter-bilayers coupling $t_{2}$, the bonding and the
antibonding electronic levels form bands whose dispersion avoids exact
compensation between the intra-bilayer coupling $t_{1\text{ }}$and the
Zeeman splitting. When $t_{1}=t_{2}$, the quasi-two-dimensional case \cite%
{bulaevski73} is retrieved: the presently studied $\pi $ phases are lost in
favor of FFLO phases modulated either along the planes or perpendicular to
the planes.

Hence we see that the interlayer coupling must be rather small to prevent
the smearing out of the electronic levels participating to the $\pi $
coupling. Extremely low coupling may in principle leads to the suppression
of the transition due to the two-dimensional fluctuations \cite{rice}.
However in the limit $t_{2}\gg T_{c}\sqrt{T_{c}/E_{F}}$ the fluctuations are
limited in a very narrow temperature region near the critical temperature 
\cite{varlamov}. For $t_{2}\ll T_{c}\sqrt{T_{c}/E_{F}}$ Kosterlitz-Thouless
regime may be attained near $T_{c}$ \cite{glazman} but the long range order
is restored outside the vicinity of the critical temperature, namely for ($%
T_{c}-T)/T_{c}>(T_{c}/E_{F})\ln (T_{c}^{3}/E_{F}t_{2}^{2}).$ This weak
logarithmic divergence of the fluctuations as $t_{2}\rightarrow 0$ means
that in practice very weak coupling is enough to restore the transition.
Finally, the FFLO-$\pi $ phase is fully established if $t_{2}\ll \Delta
_{0}^{2}/t_{1}$.

Impurities produce further broadening of the electronic levels over an
energy range $1/\tau $, $\tau $ being the elastic collision time. One may
infer that disorder leads to the destruction of field-induced
superconductivity since the $\pi $ phases originate from exact compensation
of the Zeeman splitting by the $t_{1}$ splitting between bonding and
antibonding electronic levels. As a rough estimate, $1/\tau $ plays a
similar role than the inter-bilayer coupling $t_{2}$ and thus $1/\tau =\mu
_{B}^{2}H_{0}^{2}/t_{1}$ defines a degree of disorder above which the $\pi $
phases are unlikely to survive. Consequently the observation of the above
predicted phenomena requires even cleaner samples than the observation of
the two-dimensional FFLO state which was expected to exist for $1/\tau <\mu
_{B}H_{0}$ \cite{aslamazov}.

Up to now the orbital effect has been neglected. If we choose the in-plane
coherence length as $\xi \thickapprox 100$ \AA\ and $t_{1}/E_{F}\thickapprox
10^{-2}-10^{-3}$, orbital effects are expected to be important above $%
H_{c2}^{orb}=(\phi _{0}/\xi ^{2})(E_{F}/t_{1})\thickapprox 10-100$ T, $\phi
_{0}$ being the magnetic flux quantum.

Modern nanotechnology permits now to design and fabricate a rich variety of
different superconducting structures. In the present Letter we demonstrate
that the system comprising weakly coupled superconducting bilayers may
reveal a new type of the FFLO-$\pi $ superconducting state when the spin
effect prevails against the orbital effect of the magnetic field. The
critical field of such state can exceed many times the standard paramagnetic
limit. Moreover the phenomenon of the field-induced superconductivity at low
temperature is predicted. Note that this field-induced superconductivity is
very different from that recently observed in quasi-two-dimensional organic
conductor $\lambda -(BETS)_{2}FeCl_{4}$ \cite{uji}, where the exchange field
of aligned $Fe^{3+}$ spins compensates the external field by means of the
Jaccarino-Peter effect \cite{jaccarino}.

\bigskip We are grateful to M. Houzet and A. Koshelev for useful comments.


\begin{references}
\bibitem{footnote1} Exact expressions for the Green functions will be published elsewhere
and then used to study thoroughly the low coupling $t_1 \ll \Delta_0$ limit where various 
nonuniform phases also appear, but not the presently studied  $\pi$-FFLO phase.
\end{references}
\end{document}